\documentclass[prb,aps, twocolumn, amsmath,amssymb,superscriptaddress]{revtex4}
\usepackage{float}
\usepackage{amsmath}
\usepackage{amssymb}
\usepackage{amsfonts}
\usepackage{euscript}
\usepackage{enumerate}
\usepackage{hhline}

\usepackage{pslatex}
\usepackage{tabularx}

\usepackage{graphicx}
\usepackage{dcolumn}
\usepackage{bm}
\usepackage[sort&compress]{natbib}
\usepackage{color}
\makeatletter

\renewcommand{\@biblabel}[1]{#1. }
\renewcommand{\@dotsep}{500}
\renewcommand{\@pnumwidth}{0em}
\renewcommand{\l@figure}[2]{
\@dottedtocline{1}{1.5em}{2em}{Figure #1}{}\vspace{15pt}}

\begin{document}

\title{Direct observation of nanofabrication influence on the optical properties of single self-assembled InAs/GaAs quantum dots}

\author{Jin Liu}\email{liujin23@mail.sysu.edu.cn}
\affiliation{Center for Nanoscale Science and Technology, National
Institute of Standards and Technology, Gaithersburg, MD 20899,
USA}
\affiliation{School of Physics, Sun-Yat Sen University, Guangzhou, 510275, China}
\affiliation{Maryland NanoCenter, University of Maryland, College Park, USA}

\author{Kumarasiri Konthasinghe}
\affiliation{Department of Physics, University of South Florida, Tampa, Florida 33620, USA}

\author{Marcelo Davan\c co}
\affiliation{Center for Nanoscale Science and Technology, National
Institute of Standards and Technology, Gaithersburg, MD 20899,
USA}

\author{John Lawall}
\affiliation{Physical Measurement Laboratory, National
Institute of Standards and Technology, Gaithersburg, MD 20899,
USA}

\author{Vikas Anant}
\affiliation{Photon Spot, Inc., Monrovia, CA 91016, USA}

\author{Varun Verma}
\affiliation{National Institute of Standards and Technology, Boulder, CO 80305,
USA}

\author{Richard Mirin}
\affiliation{National Institute of Standards and Technology, Boulder, CO 80305,
USA}

\author{Sae Woo Nam}
\affiliation{National Institute of Standards and Technology, Boulder, CO 80305,
USA}

\author{Jin Dong Song}
\affiliation{Center for Opto-Electronic Materials and Devices Research, Korea Institute of Science and Technology, Seoul 136-791, South Korea}

\author{Ben Ma}
\affiliation{State Key Laboratory for Superlattice and Microstructures, Institute of Semiconductors, Chinese Academy of Sciences, Beijing, 100083, China}
\affiliation{College of Materials Science and Opto-Electronic Technology, University of Chinese Academy of Sciences}
\affiliation{Synergetic Innovation Center of Quantum Information and Quantum Physics, University of Science and Technology of China, Hefei, Anhui 230026, China}

\author{Ze Sheng Chen}
\affiliation{State Key Laboratory for Superlattice and Microstructures, Institute of Semiconductors, Chinese Academy of Sciences, Beijing, 100083, China}
\affiliation{College of Materials Science and Opto-Electronic Technology, University of Chinese Academy of Sciences}
\affiliation{Synergetic Innovation Center of Quantum Information and Quantum Physics, University of Science and Technology of China, Hefei, Anhui 230026, China}

\author{Hai Qiao Ni}
\affiliation{State Key Laboratory for Superlattice and Microstructures, Institute of Semiconductors, Chinese Academy of Sciences, Beijing, 100083, China}
\affiliation{College of Materials Science and Opto-Electronic Technology, University of Chinese Academy of Sciences}
\affiliation{Synergetic Innovation Center of Quantum Information and Quantum Physics, University of Science and Technology of China, Hefei, Anhui 230026, China}

\author{Zhi Chuan Niu}
\affiliation{State Key Laboratory for Superlattice and Microstructures, Institute of Semiconductors, Chinese Academy of Sciences, Beijing, 100083, China}
\affiliation{College of Materials Science and Opto-Electronic Technology, University of Chinese Academy of Sciences}
\affiliation{Synergetic Innovation Center of Quantum Information and Quantum Physics, University of Science and Technology of China, Hefei, Anhui 230026, China}

\author{Kartik Srinivasan} \email{kartik.srinivasan@nist.gov}
\affiliation{Center for Nanoscale Science and Technology, National
Institute of Standards and Technology, Gaithersburg, MD 20899, USA}

\date{\today}

\begin{abstract}
\noindent \textbf{Single self-assembled InAs/GaAs quantum dots are a promising solid-state quantum technology, with which vacuum Rabi splitting, single-photon-level nonlinearities, and bright, pure, and indistinguishable single-photon generation having been demonstrated. For such achievements, nanofabrication is used to create structures in which the quantum dot preferentially interacts with strongly-confined optical modes. An open question is the extent to which such nanofabrication may also have an adverse influence, through the creation of traps and surface states that could induce blinking, spectral diffusion, and dephasing. Here, we use photoluminescence imaging to locate the positions of single InAs/GaAs quantum dots with respect to alignment marks with $<5$~nm uncertainty, allowing us to measure their behavior before and after fabrication.  We track the quantum dot emission linewidth and photon statistics as a function of distance from an etched surface, and find that the linewidth is significantly broadened (up to several GHz) for etched surfaces within a couple hundred nanometers of the quantum dot. However, we do not observe appreciable reduction of the quantum dot radiative efficiency due to blinking. We also show that atomic layer deposition can stabilize spectral diffusion of the quantum dot emission, and partially recover its linewidth.}
\end{abstract}

 \maketitle

\maketitle

Resonance fluorescence experiments have established that single InAs/GaAs self-assembled quantum dots (QDs) can exhibit Fourier-transform-limited emission, and as a result the individual photons emitted by these QDs can be nearly perfectly indistinguishable~\cite{Matthiesen2013,He2013,Kuhlmann2015}.  Because the collection of emission from an InAs/GaAs QD in bulk, as-grown material is limited to $<1$~$\%$ due to the total internal reflection that results from the large refractive index contrast between GaAs and air, efficient extraction of the emitted light typically requires modification of the photonic environment surrounding the QD. Such modifications should ideally not adversely influence the photon indistinguishability and recently, micropillar cavities have been able to achieve both high brightness and near-unity indistinguishability~\cite{Unsleber2016,Ding2016,Somaschi2016,He2017}. In comparison, more tightly confined geometries, such as photonic nanowires, photonic crystal cavities, and suspended waveguides, generally have not exhibited as a high degree of indistinguishability~\cite{Jons2012,Varoutsis2005,Englund2007,Madsen2014,LiuF2017,Kirsanske2017}. While this can partly be attributed to challenges in achieving high-quality resonance fluorescence in such structures (e.g., adequate suppression of the excitation laser and full control of the QD charge environment), another possibility is that the nanofabrication processes by which such structures are created may be an issue, mostly due to the plasma dry etching processes involved. In particular, fabrication of structures such as photonic crystals results in the presence of etched surfaces that are within a few hundred nanometers of the QD, and the potential influence of such surfaces on the QD emission, through coupling to surface states and charge traps, for example, is of significant concern. However, such a nanofabrication-induced effect has not been directly observed so far, in part due to the low photon extraction efficiency of QDs in bulk, inefficient single-photon detection in the 900 nm band, and challenges in tracking single QDs before and after the nanofabrication process.

To unequivocally investigate this effect, QD epitaxy with distributed Bragg reflectors and superconducting nanowire single-photon detectors (SNSPDs) optimized in the 900~nm band (quantum efficiency $>$ 80~\%) are used to enable efficient characterization of single QDs in bulk before any nanostructure fabrication. We utilize a recently-developed nanoscale optical positioning technique~\cite{Sapienza2015,Liu2017} to locate the position of those QDs with respect to alignment features with an uncertainty $<$~5 nm. Subsequent aligned electron-beam lithography and dry etching allows us to place the QDs at specified positions away from etched surfaces. By measuring the QD emission linewidth and photon statistics before and after the fabrication steps for a number of samples, we are able to directly assess the influence that nanofabrication has on these important quantities, which in part characterizes the quality of single-photon emission that is possible from these systems. We determine that etched surfaces that are within 300 nm of the QDs broaden their emission under wetting layer excitation, while blinking, though observable in systems for which the etched surfaces are 100~nm away, generally does not appreciably influence the QD radiative efficiency.  Finally, we find that the QDs that are closest to etched features have some emission lines that exhibit strong spectral diffusion, with timescales of a few seconds and a spectral wandering range of several nanometers.  Such emission lines are greatly stabilized through atomic layer deposition of an Al$_2$O$_3$ cap, which also produces a partial reduction in the homogeneous linewidth of the more stable (lower spectral diffusion) states. In agreement with recent progress in generating indistinguishable single-photons via charge stabilization techniques in p-i-n structures~\cite{Somaschi2016,Kirsanske2017}, our results suggest that control of the charge environment is likely to be necessary to achieve Fourier-transform-limited emission in nanoscale device geometries.

\vspace{-0.2in}

\section{Sample fabrication and measurement setup}
\vspace{-0.1in}

\begin{figure}[t]
\begin{center}
\includegraphics[width=\linewidth]{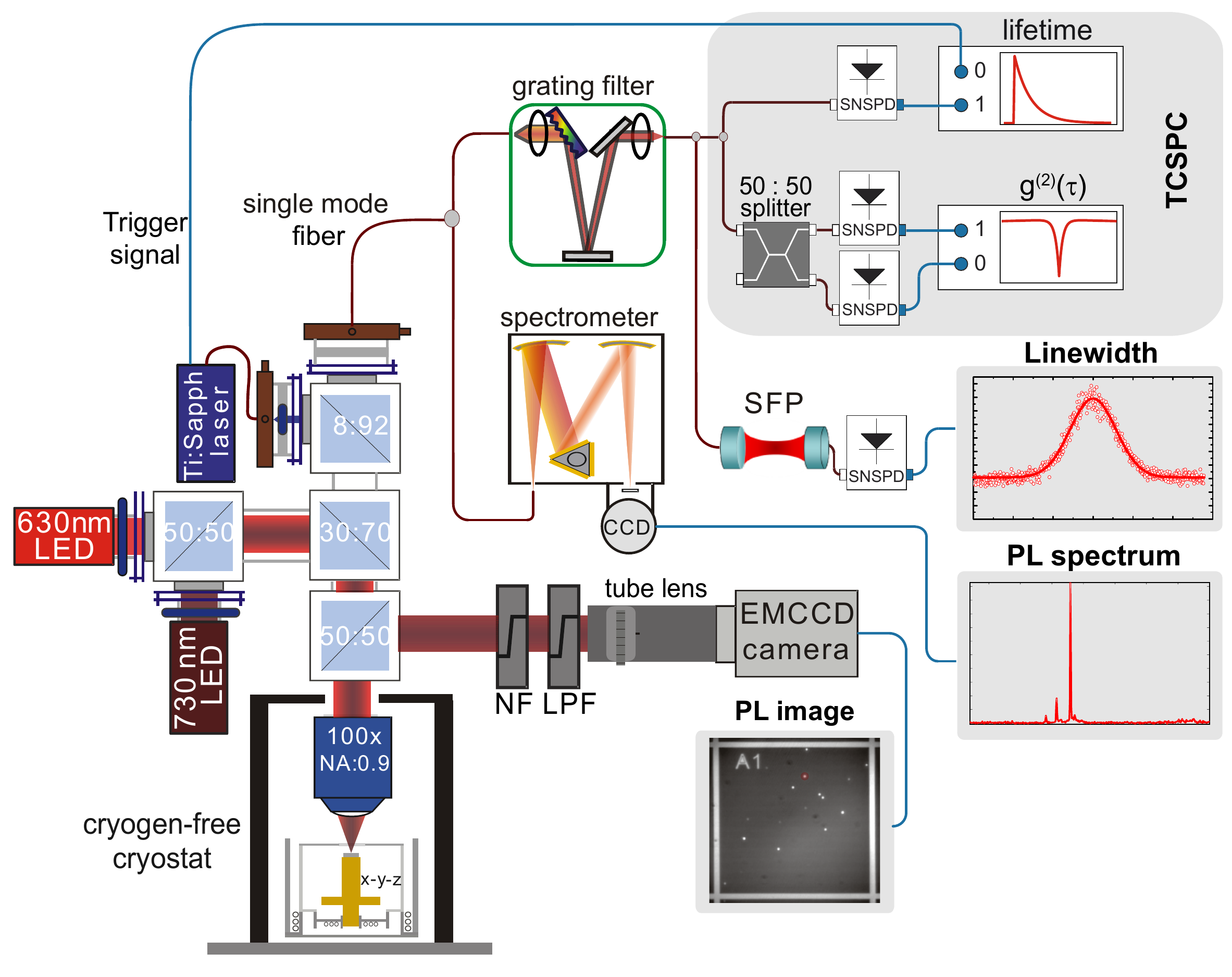}
\caption{\textbf{Experimental setup.} Photoluminescence (PL) images are generated by exciting the sample simultaneously with 630~nm and 730~nm LEDs (light emitting diodes), with PL from the QDs and reflected 730~nm light from the alignment marks separated from unwanted light through 800~nm band notch filters (NFs) and a 700~nm long-pass filter (LPF), and sent into an electron-multiplying charge-coupled device (EMCCD) camera. PL spectra from specific QDs within the image are obtained by pumping them with a fiber-coupled Ti:sapphire laser whose wavelength is tuned to the wetting layer (spot diameter $<$5~$\mu$m), collecting the emission into a single mode fiber, and sending it into a grating spectrometer. Time-correlated single-photon counting (TCSPC) measurements are performed by sending the collected emission into a fiber-coupled tunable grating filter, with the filtered signal then going into either one superconducting nanowire single photon detector (SNSPD) for lifetime measurements, or into a 50:50 fiber coupler and two SNSPDs for intensity autocorrelation measurements. Alternately, the filtered signal can be sent into a scanning Fabry-Perot (SFP) cavity for high-resolution spectroscopy, where the output of the SFP is again coupled into a SNSPD.}
\vspace{-0.3in}
\label{fig:Fig1}
\end{center}
\end{figure}

We consider two circularly symmetric geometries to assess the influence of nanoscale etched features on QD behavior.  The first is the suspended circular Bragg grating geometry previously studied in Refs.~\onlinecite{Davanco2011,Sapienza2015}.  This geometry, which consists of a central circular region (diameter of 1.2~$\mu$m) surrounded by etched circular grooves, can improve the radiative properties of the QD, through coupling to a localized cavity mode that enables Purcell enhancement of the QD radiative rate and preferential upwards emission with a relatively narrow divergence angle.  The potential adverse effects of this geometry, in terms of an increased QD emission linewidth (e.g., due to spectral diffusion or additional dephasing channels) or the introduction of additional dark states, has not yet been systematically investigated.  While a previous study~\cite{Davanco2014} indicated that multiple timescale blinking of QDs in these cavities could result in a significant decrease in the radiative efficiency below unity, the QDs were not studied prior to device fabrication, so that it was not possible to assess the role that etched surfaces played in the creation of the dark states that induced blinking.

The second structure is a nanopillar with a circular cross-section.  This is chosen as a convenient means to place etched surfaces within the vicinity of a QD, and we vary the pillar diameter across devices from 100~nm to 600~nm, so that the maximum distance the QD can be from an etched surface is between 50~nm to 300~nm.  The QD epitaxy (see supplementary information, S.I.) contains a layer of InAs QDs embedded within a 160~nm thick GaAs layer (80~nm below the surface), and makes use of an underlying distributed Bragg reflector (DBR) stack to enhance the upwards vertical emission for the QDs in bulk and nanopillar geometries.  It also contains a 1000~nm thick Al$_{0.7}$Ga$_{0.3}$As sacrificial layer, placed in between the QD-containing layer and the DBR, which allows the aforementioned suspended grating microcavities to be fabricated from the same epitaxy, so that the growth conditions are fixed across all samples.

After fabricating metallic alignment marks on the sample (see S.I.), we use our recently-developed photoluminescence imaging system~\cite{Liu2017,He2017}, shown schematically in Fig.~\ref{fig:Fig1}, to locate the positions of single QDs with a one standard deviation uncertainty $<$5~nm.  This technique relies on wide-field excitation of QDs within an $\approx66$~$\mu$m~$\times$~66~$\mu$m field of view using a short-wavelength LED, and simultaneous illumination of the sample surface (including metallic alignment marks) using a second, longer-wavelength LED. Emission from the QDs and reflected illumination light are separated from unwanted light using filters, and sent to a sensitive camera, where the generated image is analyzed using a maximum likelihood estimator and a cross-correlation approach for identifying the centers of the QD emission and alignment marks, respectively.  The QD locations are used in subsequent device fabrication, which consists of aligned electron-beam lithography, plasma etching of the QD-containing GaAs layer, resist removal, and in the case of the circular grating cavities, a hydrofluoric acid etch to remove the Al$_{0.7}$Ga$_{0.3}$As sacrificial layer.

\begin{figure*}
\begin{center}
\includegraphics[width=0.75\linewidth]{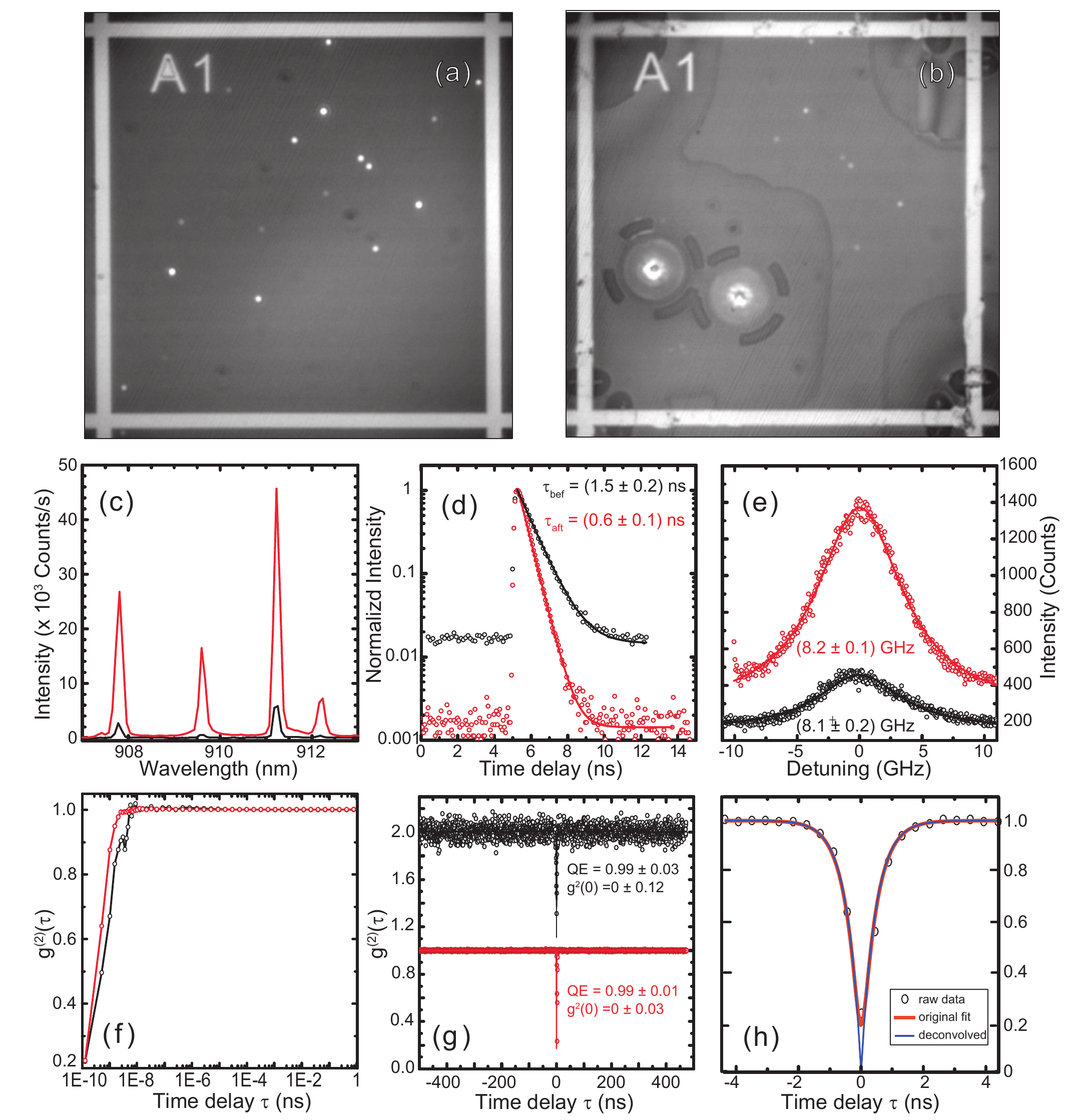}
\caption{\textbf{QDs in circular Bragg grating cavities}. The optical properties of QDs before and after the creation of circular Bragg gratings are measured.  (a)-(b) Photoluminescence images of the QDs (a) before and (b) after device fabrication. The alignment mark separation is 50~$\mu$m. (c) Photoluminescence spectrum, (d) photoluminescence decay, (e) emission linewidth, and (f) - (h) intensity autocorrelation recorded for one of the QDs before (black lines/circles) and after (red lines/circles) fabrication, under wetting layer excitation. The data in (d)-(h) is taken for the brightest emission line in the spectrum from (c) (near 911~nm). The photoluminescence decay data in (d) is fit to a monoexponentially decaying function, and a lifetime $\tau_{\text{bef}}=1.5$~ns~$\pm$~0.2~ns ($\tau_{\text{after}}=0.6$~ns~$\pm$~0.1~ns) is extracted for the QD state before (after) fabrication, where the uncertainty is a one standard deviation value from the fit. The emission linewidth data (circles) in (e), measured by the SFP cavity, is fit to a Gaussian function (solid line) to determine the full-width at half-maximum listed on the plot, with the uncertainty being a one standard deviation value from the fit. The intensity autocorrelation measurement in (f) was recorded over a duration long enough to enable evaluation of $g^{(2)}(\tau)$ out to a time delay as long as 1~s, and the time-axis is given in a logarithmic scale.  The intensity autocorrelation measurements in (g)-(h) are presented over a narrower range of time delays, to focus on the antibunching dip and potential presence of shorter timescale blinking. For both the before- and after-fabrication data in (g), the quantum efficiency (QE) is extracted from a three-level system fit to the data, as described in the main text, and the uncertainty value is a one standard deviation value from the fit.  The quoted $g^{(2)}(0)$ values are determined by additionally deconvolving the SNSPD/TSCPC timing response from the fit. A zoomed-in view of the raw data, fit (no deconvolution), and fit including deconvolution is shown in (h), for the QD state after device fabrication.  The before-fabrication data in (g) is vertically shifted up by 1.0 units for clarity.}
\label{fig:Fig2}
\vspace{-0.3in}
\end{center}
\end{figure*}

Our photoluminescence imaging system also contains a confocal path for excitation of individual QDs using a continuous-wave Ti:sapphire laser, and collection of QD emission into a single mode fiber. The QDs are all excited at the wetting layer transition wavelength ($\approx$~850~nm), as successful excitation through lower energy states (e.g., quasi-resonant excitation via the QD p-shell or resonant excitation of the QD transition) was not consistently observed for all QDs. Collected emission is either sent into a grating spectrometer for spectral analysis, or through a tunable grating filter for spectral isolation of individual QD transitions.  Isolated QD lines are then sent into a scanning Fabry-Perot (SFP) cavity for high-resolution (200~MHz) linewidth analysis, or through a 50:50 fiber-coupled beamsplitter and into two SNSPDs and a time-correlated single-photon counting (TCSPC) card for measurement of the intensity autocorrelation function ($g^{(2)}(\tau)$).  Finally, a pulsed excitation source is used for measurements of the radiative decay of a given QD transition.  All measurements are performed on the same QD both before and after device fabrication.

\vspace{-0.1in}
\section{Circular Bragg grating devices}
\vspace{-0.1in}

Figure~\ref{fig:Fig2} shows representative results for the circular Bragg grating cavities.  QDs within field A1 are located using the aforementioned imaging approach (Fig.~\ref{fig:Fig2}(a)), and circular Bragg grating cavities are fabricated around two of the located QDs (Fig.~\ref{fig:Fig2}(b)).  The effects of the microcavity on the QD radiative properties are clearly seen in Fig.~\ref{fig:Fig2}(c)-(d), and consist of a strong increase in the collected emission under saturated excitation (Fig.~\ref{fig:Fig2}(c)), and an $\approx$2.5$\times$ reduction in the QD radiative lifetime of the 911~nm transition line (Fig.~\ref{fig:Fig2}(d)). Considering the $\approx$3.6~nm detuning between the QD and cavity mode line center, this level of Purcell enhancement is consistent with the QD being spatially offset from the center of the device by no more than 50~nm~\cite{Sapienza2015}. Encouragingly, we also do not observe any adverse influence of fabrication, as evidenced by measurements of the linewidth (Fig.~\ref{fig:Fig2}(e)) and the intensity autocorrelation function ($g^{(2)}(\tau)$) of this QD transition (Fig.~\ref{fig:Fig2}(f)-(h)).

In the $g^{(2)}(\tau)$ measurement, we have recorded data over a duration long enough to enable its evaluation over 10 orders of magnitude in time (Fig.~\ref{fig:Fig2}(f)).  As discussed in Ref.~\onlinecite{Davanco2014} in the context of InAs/GaAs QDs, and in several earlier works focused on the behavior of single molecules and colloidal QDs~\cite{Orrit2005}, measurement of $g^{(2)}(\tau)$ out to sufficiently long timescales can be a preferred approach for studying blinking in single quantum emitters.  In particular, while a time record of the fluorescence intensity is sensitive to the time bin width chosen (photon shot noise dominates for too small bins; shorter timescale behavior is washed out for too long bins), and subsequent histogramming analysis is influenced by the choice of a threshold intensity level, $g^{(2)}(\tau)$ does not require selection of such potentially arbitrary input parameters\cite{Note}. Blinking is evidenced in the $g^{(2)}(\tau)$ data by bunching ($g^{(2)}>1$) after the initial anti-bunching dip at $\tau=0$, and the subsequent transition to the Poissonian level ($g^{(2)}(\tau)=1$) can occur over second-long timescales, and is potentially punctuated by several steps. In Ref.~\onlinecite{Davanco2014}, such behavior was well-reproduced by a model in which the radiative transition is coupled to multiple dark states, with each dark state responsible for a step in $g^{(2)}(\tau)$, and showing a characteristic occupancy and population and de-population rate.

Here, we see no pronounced bunching or multiple dark state behavior, in either the before- or after- fabrication data in Fig.~\ref{fig:Fig2}(f).  Because $g^{(2)}(\tau)=1$ for $\tau>10$~ns (in contrast, in Ref.\onlinecite{Davanco2014}, the Poissonian level was reached only at $\mu$s or even 100~ms timescales), we zoom in on the region within $\pm~500$~ns of $\tau=0$ in Fig.~\ref{fig:Fig2}(g), and analyze the data by fitting it to three-level model in which the radiative transition is coupled to a single dark state (see supplementary material).

The SFP-measured homogeneous linewidth is essentially unchanged at $\approx$~8~GHz while the enhancement of the single-photon extraction efficiency is clearly seen from the photon counts in the SFP measurement. In the circular Bragg grating geometries, both radiative decay rates and single-photon collection efficiency of single QDs have been significantly enhanced by coupling to the confined cavity mode while other crucial properties (i.e., linewidth and quantum efficiency) are unchanged, which is highly desirable for realizing bright and coherent single-photon sources for quantum information processing tasks.

\section{Nanopillar devices}

\begin{figure*}
\begin{center}
\includegraphics[width=\linewidth]{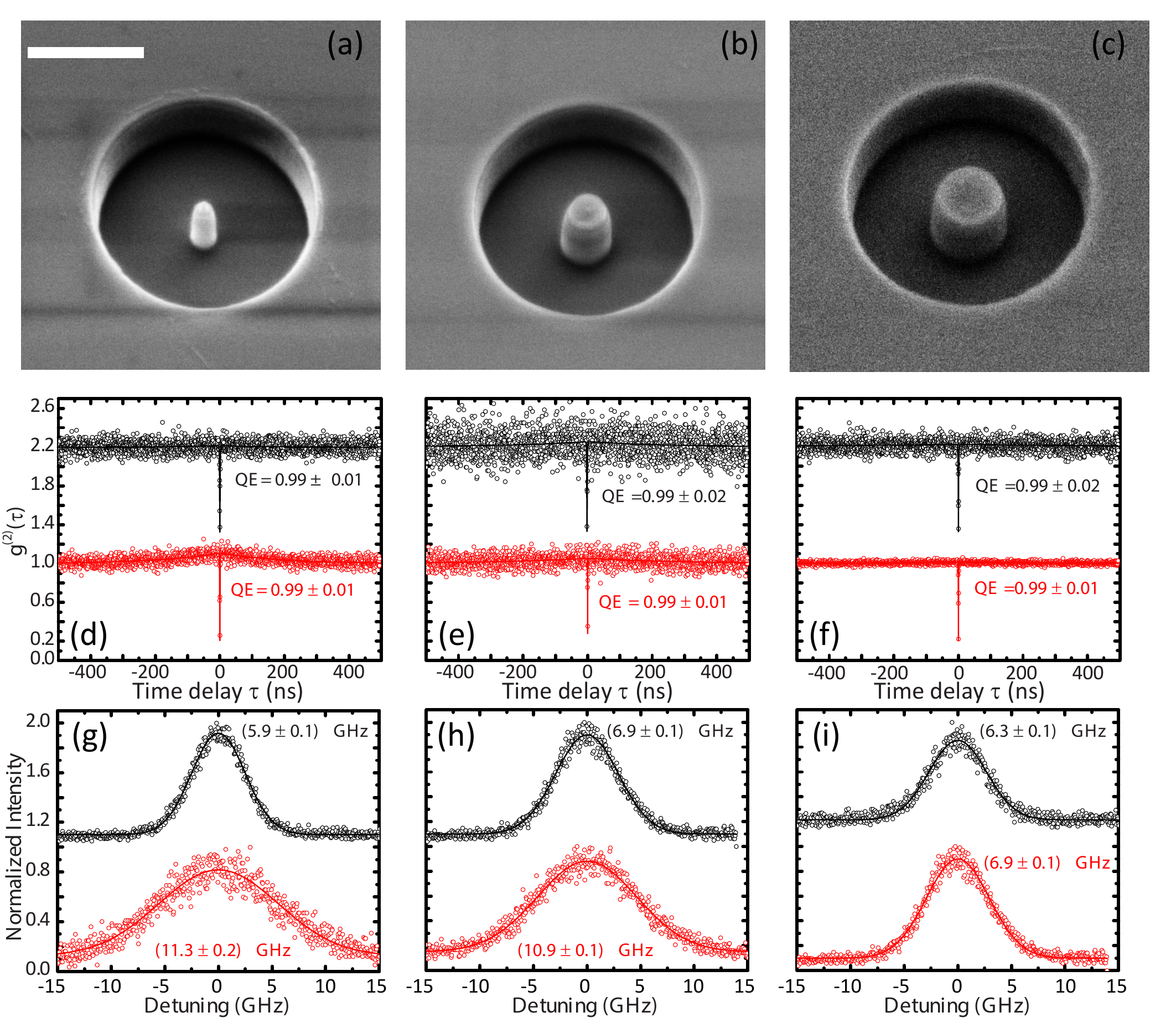}
\caption{\textbf{Behavior of QDs in etched nanopillars}. Behavior of QDs before (black lines/circles) and after (red lines/circles) fabrication of nanopillars of varying diameter, under wetting layer excitation. (a) Scanning electron microscope (SEM) image, (d) intensity autocorrelation, and (g) emission linewidth for a QD in the smallest nanopillar fabricated, with a nominal diameter of 100~nm. (b) SEM image, (e), intensity autocorrelation, and (h) emission linewidth for a QD in a nanopillar with a nominal diameter of 300~nm. (c) SEM image, (g), intensity autocorrelation, and (i) emission linewidth for a QD in a nanopillar with a nominal diameter of 600~nm.  The before-fabrication data in (d)-(f) is vertically shifted up by 1.2 units for clarity. The QE values in (d)-(f) are extracted from a three-level system fit to the data (solid lines), as described in the main text, and the uncertainty value is a one standard deviation value from the fit. The emission linewidth data in (g)-(i) are fit to Gaussians  to determine the full-width at half-maximum listed on the plots, with the uncertainties being one standard deviation values from the fits. The scale bar displayed in (a) represents 1000~nm and is applicable to the SEM images in (b) and (c) as well.}
\label{fig:Fig3}
\end{center}
\end{figure*}

The linewidth broadening effect induced by smooth epitaxial interfaces parallel to the sample surface has previously been studied in detail with superlattice structures~\cite{Wang2004,Houel2012}. In our study, the combination of high-accuracy QD positioning and high-resolution linewidth measurements allows us to investigate influences from the etched sidewalls that are ubiquitous in planar nanophotonic devices. Such dry etched surfaces usually experience a combination of strong physical ion bombardment and complicated chemical reaction processes, and are potentially more likely to introduce surface traps/states than the smooth epitaxial interfaces.

In a circular Bragg grating cavity containing a single, accurately positioned QD, the nearest etched surface is $\approx$ 600~nm away from the QD. Such a relatively large separation ensures that the confined exciton states in the QD are immune to the influence of any surface traps/states created by the dry etching process. On the other hand, more tightly confined optical modes with a sub-cubic-wavelength scale volume are highly desirable for achieving stronger light-matter interaction, e.g, the strong coupling regime with its accompanying single-photon-level nonlinearity. In such small mode volume nanophotonic structures, such as photonic crystal cavities, it is often inevitable that the optimal position for QD-field interaction will be within the vicinity (few hundred nm) of etched surfaces. Thus, we fabricate QD-containing nano-pillars with different diameters, to further investigate the influences of etched surfaces on QDs that are nominally 50~nm, 100~nm, and 300~nm away from the dry etched sidewalls, shown in Fig.~\ref{fig:Fig3}(a-c). Since there is neither an engineered cavity resonance nor far-field reshaping effect, we don't expect any pronounced Purcell effect and collection efficiency enhancement in these nano-pillars. Again, we focus on the optical properties (i.e., blinking and linewidth) that are crucial for single-photon generation and potentially influenced by the presence of the etched surfaces.

For the QD that is nominally 50~nm away from the etched surface (Fig.~\ref{fig:Fig3}(a)), a very small bunching peak near zero time delay is observed in the $g^{(2)}(\tau)$ measurement, which is a signature of coupling to the dark states (Fig.~\ref{fig:Fig3}(d)). By fitting the long-time scale $g^{(2)}(\tau)$ with the three-level-system model (see S.I.), we extract a quantum efficiency of $0.99\pm0.01$, which is nearly unchanged compared to the value before-fabrication, indicating that the coupling to any dark states that is induced by the etched surfaces is too small to appreciably change the quantum efficiency of the QD. This is strikingly different from the case in Ref.\onlinecite{Davanco2014}, in which coupling to dark states lowered the quantum efficiency of the QD down to 78~\% (see the comparison of QE between Ref.\onlinecite{Davanco2014} and this work in the S.I.). We note that in our previous work, in which the QD was not characterized pre-fabrication\cite{Davanco2014}, it was not possible to determine whether blinking was an intrinsic property of the as-grown QD, e.g., potentially due to introduction of defects during growth, or whether it was induced by either sample annealing~\cite{Malik1997} pre-etching, or by the etch process. The present data strongly indicates that etching likely did not play a big role.

The linewidth of the QD that is 50~nm away from the etched surface is more sensitive than the quantum efficiency and broadened by a factor of $\approx$~1.9 (Fig.~\ref{fig:Fig3}(g)). We postulate that such a linewidth broadening is mainly due to the spectral diffusion induced by the charge states on the etched surfaces. By moving the QD to nominally 150~nm away from the surface (Fig.~\ref{fig:Fig3}(b)), the bunching near the zero delay in the $g^{(2)}(\tau)$ curve is almost negligible, resulting in an unchanged quantum efficiency after the fabrication. The linewidth broadening factor is reduced to 1.72, indicating an alleviated influence from the etched surface. Once the QD is nominally 300~nm away from the etched surface (Fig.~\ref{fig:Fig3}(c)), we barely observe any changes either in $g^{(2)}(\tau)$ or in the linewidth measurement (Figs.~\ref{fig:Fig3}(f) and ~\ref{fig:Fig3}(i), respectively).

\begin{figure}
\begin{center}
\includegraphics[width=\linewidth]{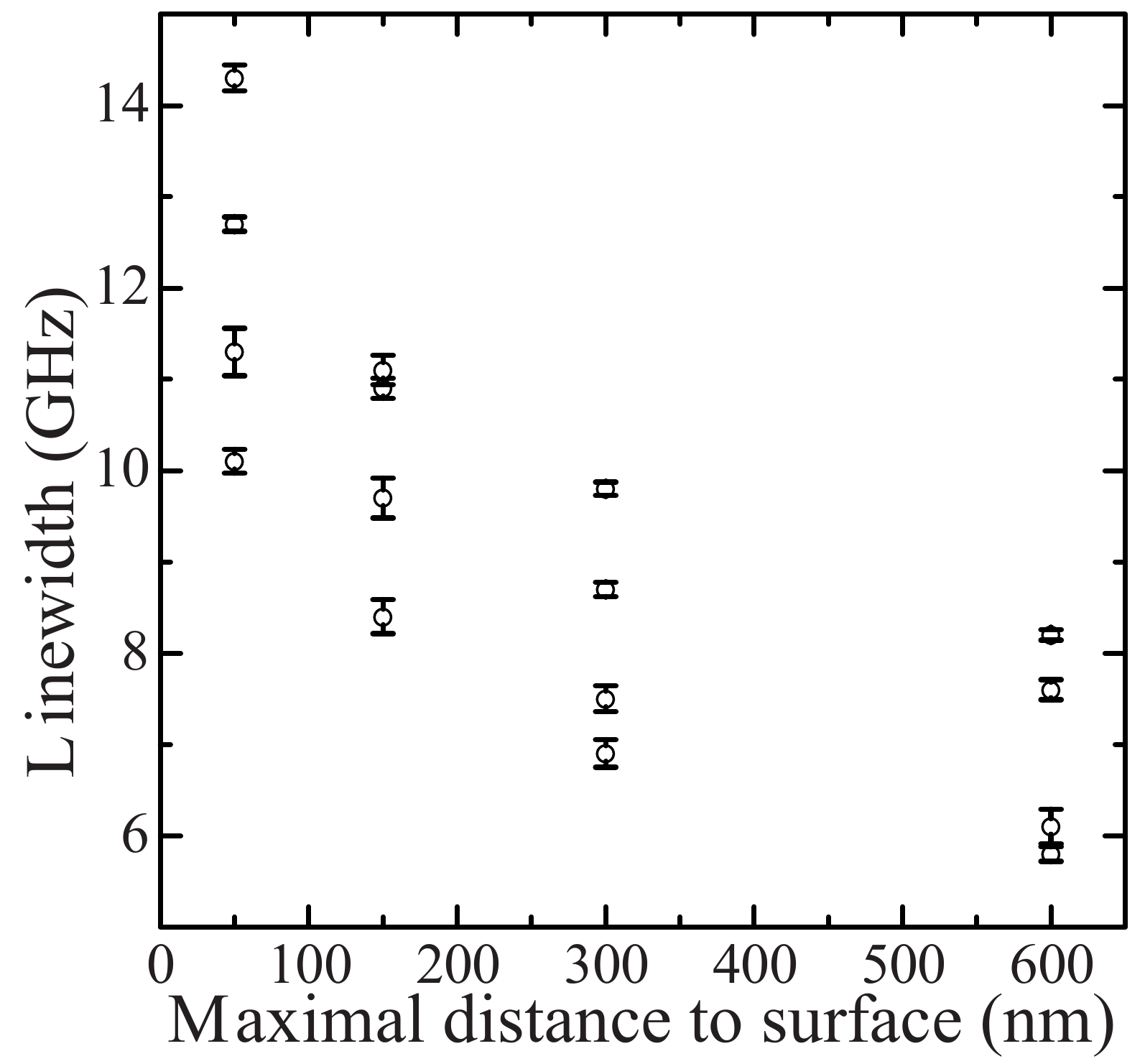}
\caption{\textbf{Linewidth of QDs as a function of proximity to etched surfaces}. Compilation of QD emission linewidths under wetting layer excitation as a function of the maximum distance between the QD and an etched surface.  For the circularly symmetric structures studied in this work, this is given by the radius of the central region of the circular Bragg grating (Figs.~\ref{fig:Fig2}) and the radius of the nanopillar 80~nm below the top surface (Fig.~\ref{fig:Fig3}).  The emission linewidth is measured by the SFP method, and the error bars are given by the one standard deviation uncertainties from nonlinear least squares fits of the data to Gaussian functions.}
\label{fig:Fig4}
\end{center}
\end{figure}

Although it is very informative to measure the linewidth of the same QDs before and after the fabrication, the very long characterization time of these measurements prevents us from obtaining a large sample of data for additional statistical analysis. Thus, we have positioned a number of QDs in nano-pillars with various sizes without systematic optical characterization prior to the fabrication. Fig.~\ref{fig:Fig4} presents the statistics of the linewidth of QDs with different maximal distances to the etched surfaces. We clearly see the linewidth broadening effect is significantly reduced by moving the QDs away from the etched surface. The critical distance to avoid the surface charges is approximately 300~nm, for which the linewidth of QDs is close to the number for bulk QDs. We note that average linewidth of the QDs in bulk is $\approx$ 6 GHz under the wetting layer excitation scheme used in this work. A more sensitive probe of the charge environment could be achieved by using a resonant excitation scheme~\cite{Houel2012}; however the QDs used in this study did not exhibit resonance fluorescence signals.

\section{Surface passivation via atomic layer deposition}
\begin{figure*}
\begin{center}
\includegraphics[width=\linewidth]{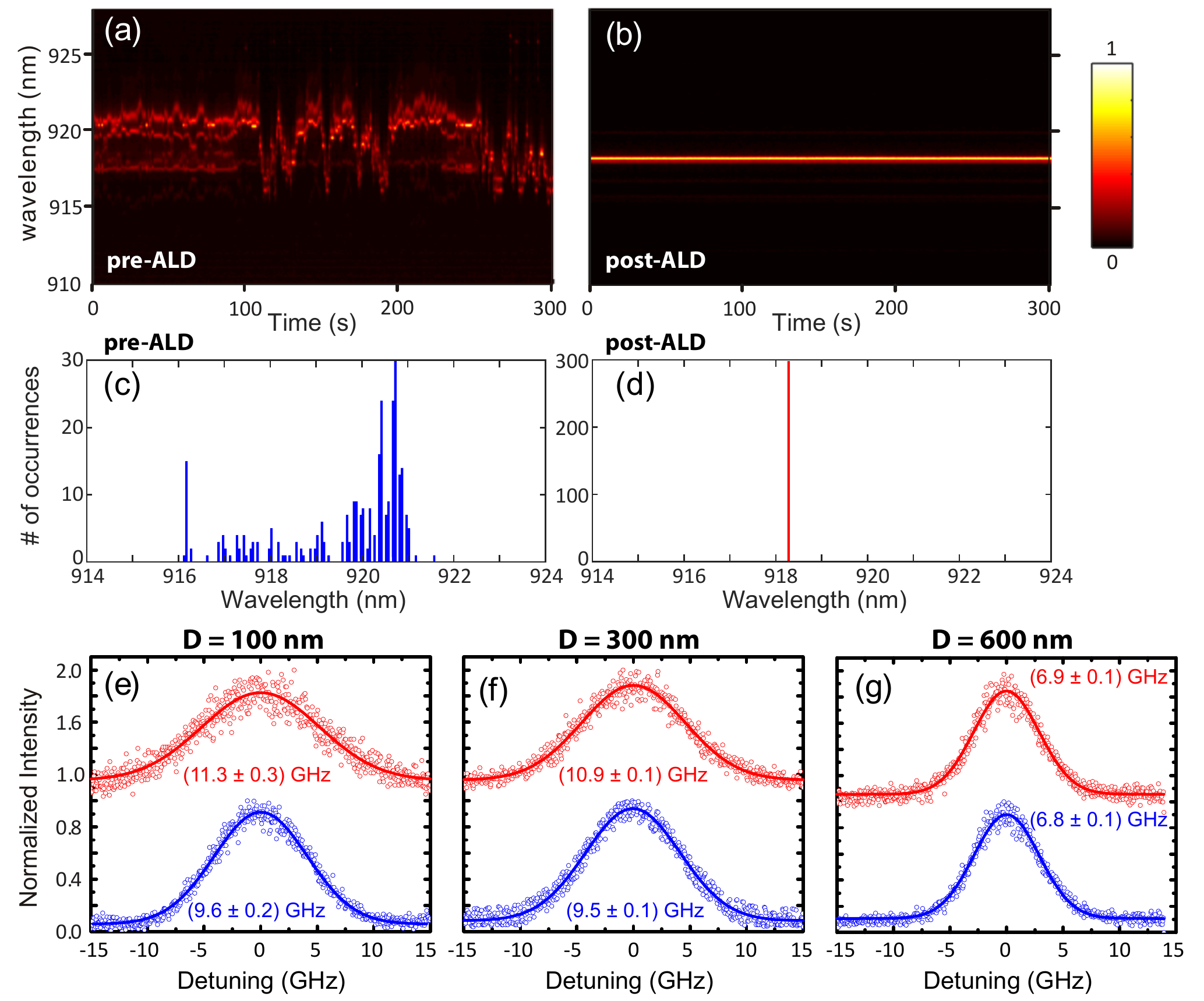}
\caption{\textbf{Stabilizing QD emission through atomic layer deposition}. Photoluminescence spectrum, recorded as a function of time in 1~s acquisition intervals, for a QD (a) before and (b) after atomic layer deposition (ALD) of an Al$_2$O$_3$ capping layer. In each case, the emission intensity is normalized to the maximum value within the spectrum (scale bar shown to the right). (c),(d) are the histograms of the photon counts as a function of wavelength in (a) and (b) respectively, with a spectral bin size of 0.05~nm. (e)-(g) Emission linewidth for QDs in nanopillars before (red lines/circles) and after (blue lines/circles) ALD.  The nanopillar diameters are (e) 100~nm, (f) 300~nm, and (g) 600~nm. The emission linewidth is measured by the SFP method, and the uncertainties are given by the one standard deviation uncertainties from nonlinear least squares fits of the data to Gaussian functions.}
\label{fig:Fig5}
\end{center}
\end{figure*}

Due to the uncertainties from the e-beam lithography and the optical positioning process, we could occasionally achieve a situation where the QDs are less than 50 nm away from the etched surfaces. These QDs typically exhibit severe spectral wandering on a time scale of a few seconds~\cite{Robinson2000}, as shown in Fig.~\ref{fig:Fig5}(a). The single QD shows two typical spectral characteristics: several emission lines with slightly varying wavelength, and a single emission line that jumps across a range of a few nanometers. Since surface treatments have been demonstrated as very effective ways to improve emission and reduce absorption in III-V material~\cite{YEO2011,Favero2017}, we employ an atomic layer deposition (ALD) process to deposit a thin capping layer of Al$_{2}$O$_{3}$ for stabilizing the charge environment on the etched surfaces. After depositing a 15~nm Al$_{2}$O$_{3}$ capping layer, the large spectral wandering effect is completely suppressed and the single QD emission line is stabilized with an enhanced emission intensity, as shown in Fig.~\ref{fig:Fig5}(b). The suppression of spectral wandering can be further appreciated in Fig.~\ref{fig:Fig5}(c,d) showing the histogram of the QD's center emission wavelength in Fig.~\ref{fig:Fig5}(a,b) respectively. Prior to the ALD process, the higher counts near to 916~nm and 920~nm correspond to the multi-exciton complex states with several emission lines, while the low counts in between reveal the wavelength shifting of the single exciton state, shown in Fig.~\ref{fig:Fig5}(c). The stable single emission line due to the effective removal of the surface charges by ALD results in a $\delta$-function-like histogram in Fig.~\ref{fig:Fig5}(d).

We further apply the ALD process to the nanopillar devices from the previous section (which did not exhibit large spectral wandering) and investigate the surface passivation effect on the QD linewidth. Figure~\ref{fig:Fig5}(e-g) presents the linewidth of the QDs shown in Fig.~\ref{fig:Fig3}(g-i), with and without the ALD process. The linewidth of the QDs in the nanopillar devices is indeed reduced; however, it does not fully recover to the value prior to dry etching. Further linewidth reduction is not observed with an extra deposition of 15~nm Al$_{2}$O$_{3}$, indicating that the charge stabilization induced by Al$_{2}$O$_{3}$ has been fully established with only one ALD step. The stabilization of the QD emission lines and partially recovered linewidth by the ALD process suggest that surface passivation processes could be a viable method to remedy the adverse effects introduced by the presence of etched surfaces that results from the fabrication of nanophotonic devices.

\vspace{0.1in}
\section{Discussion}
In conclusion, we have directly and quantitatively investigated the influence of nanofabrication on the optical properties of single QDs via a fluorescence-imaging-based QD positioning technique, thereby shedding light on considerations that must be taken into account when designing and building high-performance QD-based quantum photonic devices via modern nanofabrication processes. First, by positioning QDs in the center of circular Bragg grating cavities, $\approx$ 600 nm away from etched surfaces, we find that we can simultaneously increase both the decay rate and collection efficiency of the QD, without compromising its optical properties.  We then consider nanopillar geometries in which the separation of the QD from the etched surfaces is reduced.  While the quantum efficiency of single QDs is rather insensitive to the surfaces, the linewidth starts to broaden once the QD is within 300 nm of the surface, at which point charged surface states appear to play an important role. Strong spectral wandering is observed when the distances between the QDs and surfaces are less than 50~nm. An ALD process is successfully applied to completely suppress the strong spectral wandering and partially reduce the linewidth broadening.

In the future, higher-resolution spectroscopy, e.g., based on resonance fluorescence, is highly desirable to fully probe the adverse effects from the nanofabrication process. Such measurements could better elucidate the role of spectral wandering on the observed linewidth broadening~\cite{Houel2012}.  Controlling such spectral wandering will likely require some form of charge stabilization, as has recently been successfully applied by a number of groups through the use of p-i-n structures to enable high-performance single-photon sources~\cite{Kuhlmann2015,Somaschi2016,Kirsanske2017}. Further implementation of such p-i-n structures in nanophotonic devices with ultra-small mode volumes may impose formidable technical challenges on the fabrication process. Alternatively, surface passivation could serve as a potentially important approach to stabilize the charge environment without introducing designs that require electrical contacts. Thus, more advanced surface passivation techniques than the ALD approach presented here could be pursued to maximally limit the influences of charged states from the etched surfaces.

\noindent \textbf{Acknowledgements}
J.L. acknowledges support under the Ministry of Science and Technology of China (grant no. 2016YFA0301300), the National Natural Science Foundation of China (grant no. 11304102) and under the Cooperative Research Agreement between the University of Maryland and NIST-CNST, Award 70NANB10H193. Z.C.N acknowledges support under National Key Basic Research Program of China (2013CB933304), and the National Natural Science Foundation of China (91321313).

\noindent \textbf{Author Contributions}
J.L. built the optical setup and performed the measurements together with K.K. and M.D., K.K and J.La built the scanning Fabry-Perot interferometer, V.A., V.V., R.M. and S.W.N. developed the SNSPD system. B.M.,J.D.S.,Z.S.C.,H.Q.N. and Z.C.N. performed the epitaxial growth. The project was initialized and supervised by K.S. J.L, M.D, and K.S analyzed the data and wrote the manuscript.



\newpage
\onecolumngrid \bigskip

\begin{center} {{\bf \large SUPPLEMENTARY MATERIAL}}\end{center}

\setcounter{figure}{0}
\makeatletter
\renewcommand{\thefigure}{S\@arabic\c@figure}

\setcounter{equation}{0}
\makeatletter
\renewcommand{\theequation}{S\@arabic\c@equation}

\section{sample layer structure}
The epitaxial structure consists of a 190~nm thick GaAs slab incorporating a layer of self-assembled InAs/GaAs quantum dots at the center, grown on a 1000~nm thick $\mathrm{Al_{0.7}Ga_{0.3}As/GaAs}$ sacrificial layer. Under the sacrificial layer, a distributed-Bragg-reflector with 24 pairs of $\mathrm{Al_{0.9}Ga_{0.1}As/GaAs}$ $\mathrm{\lambda/4}$ layers was gown to improve the light extraction efficiency, see Fig.~S1.

\begin{figure}[htbp]
\includegraphics[width=0.8400\textwidth,angle=0]{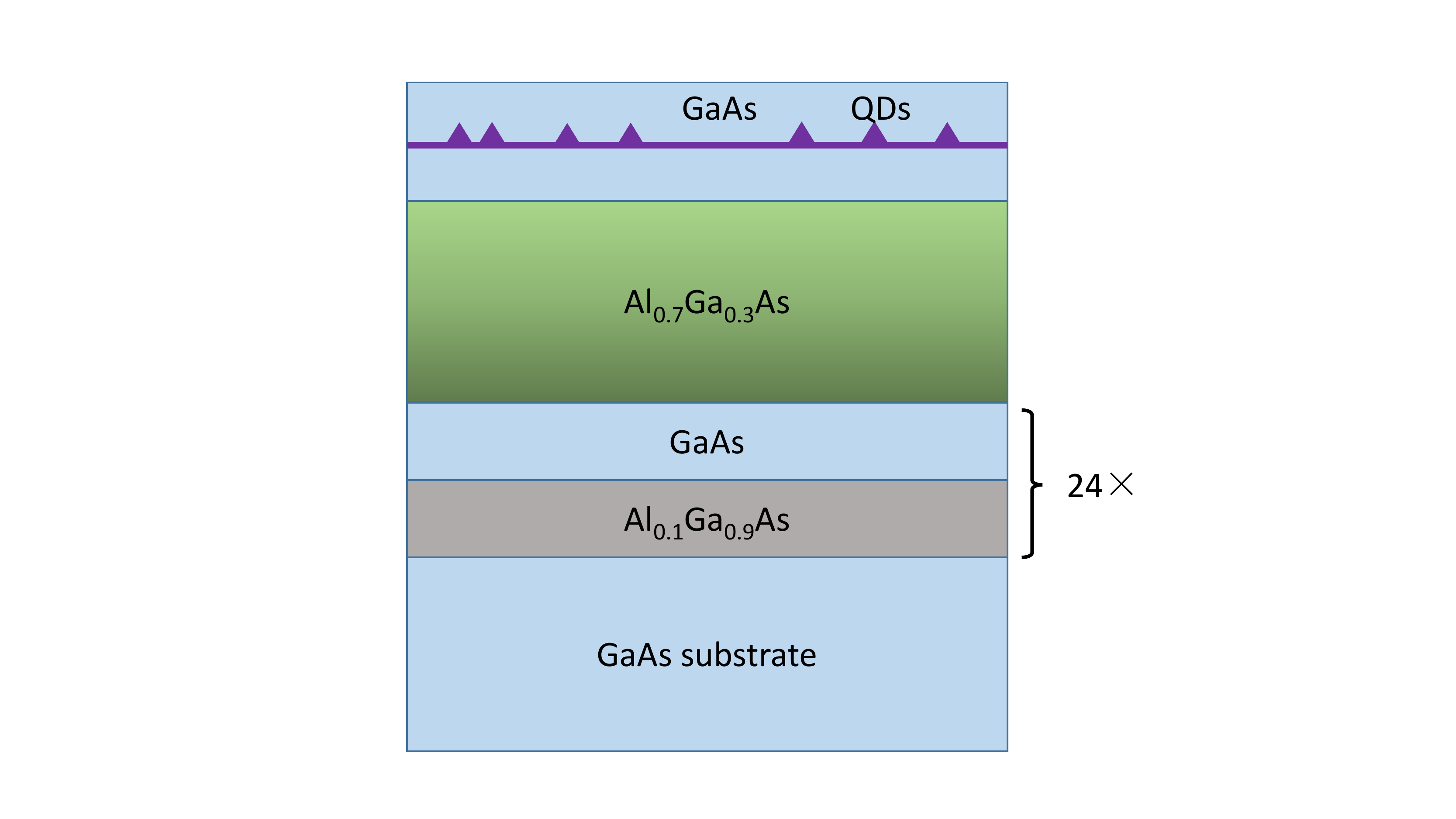}
\caption{\label{Fig1_s1.} Sample structure.}
\end{figure}
\section{Fabrication Details}
The fabrication process is similar to our previous work in Ref.~\onlinecite{Sapienza2015}, in which metallic alignment marks are firstly defined with electron-beam lithography (EBL) and lift-off processes. After the quantum dot positions are extracted via the fluorescence imaging technique, a second EBL and a chlorine-based dry etch process are applied to transfer the designed patterns (circular Bragg grating and nano-pillar) into the GaAs layer. For the circular Bragg gratings, a hydrofluoric acid undercut is required to form suspended structures.

The nano-pillar devices are treated by two atomic layer deposition (ALD) processes to form thin/uniform Al$_{2}$O$_{3}$ films on the surface of the nano-pillars. Our ALD process, implemented with the FlexAL systems from Oxford instruments~\cite{NIST}, consists of serial pulses cycled in a rapid succession: trimethyl-aluminium (TMA)chemisorbtion, TMA purge, O2 plasma and post plasma purge. The depositions are performed at 300 C. For each deposition process, we run 150 precursor cycles with a growth rate $\approx$ 0.1 nm per cycle for alumina.

\section{Second-order correlation measurements and quantum efficiency estimation}
A Hanbury-Brown and Twiss (HBT) setup was used to obtain the second-order correlation function $g^{(2)}(\tau)$ of the QD emission upon continuous-wave pumping at the saturation power. We recorded single photon detection events with a time correlator in time-tagged mode for 2~h or 4~h, depending on the count rates in the SNSPDs. We fit the $g^{(2)}(\tau)$ data, including the convolution of the detector response, by using a rate equation model in which a bright exciton transition is coupled to multiple dark states, as shown in Fig.~S2. The populations $p$ of each state evolves according to the rate equations:
\\
\begin{eqnarray}
\frac{dp_{\rm X_B}}{d\tau} &=&
r_{\rm up} p_{\rm G}-r_{\rm down} p_{\rm X_B}\nonumber
\\
\frac{dp_{\rm G}}{d\tau} &=&
r_{\rm down} p_{\rm X_B} + \sum_i d_i p_i-\sum_i u_i p_{\rm X_G}\nonumber
\\
\frac{dp_{i}}{d\tau} &=& u_i p_{\rm X_G}-d_i p_i\nonumber
\end{eqnarray}

In such a model, each dark state is populated at a rate $u_{i}$ and de-populated at a rate $d_{i}$, and all parameters are varied in the fit except for the radiative decay rate, $r_{down}$, which is determined from independent lifetime measurements. The optical transitions measured in this work are all single-exciton states, revealed by the power dependent fluorescence measurements, and the $g^{(2)}(\tau)$ data can be very well fitted with only one dark state. The quantum efficiencies of the QDs then are extracted from the estimation of the dark state occupancy.

\begin{figure}[htbp]
\includegraphics[width=0.8400\textwidth,angle=0]{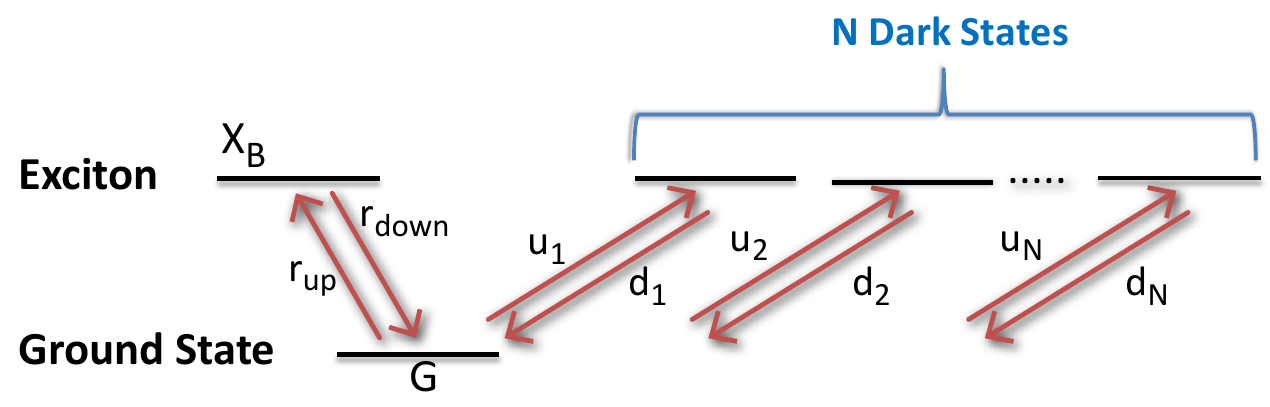}
\caption{\label{Fig2_s2} \textbf{Energy level of the QD system coupling to the dark state.} $X_{B}$/$X_{B}$ and G are the optically bright/dark and ground states of the exciton respectively. Pumped by the from the ground state G with a rate of $r_{up}$, the bright state $X_{B}$ decay to the ground state with a spontaneous decay rate $r_{down}$.}
\end{figure}

\section{Comparison of the quantum efficiency between 100~nm diameter nanopillar and device 1 in Ref.~S3}
We plot in Fig.~S3 the long time scale $g^{(2)}(\tau)$ data of the 100~nm nanopillar and device 1 in Ref.~\onlinecite{Davanco2014} together to directly compare different blinking behaviors. For the QD in Ref.~\onlinecite{Davanco2014}, the anti-bunching level right after the zero time delay is up to 1.3 and slowly decays, with a few discrete slope changes, towards the Poissonian level at $\approx$ 1~s. Such a behavior is well fitted by using the rate equation model with 7 dark states in Ref.~\onlinecite{Davanco2014}, which gives a quantum efficiency of $\approx$~78~\%. In contrast, the magnitude of the anti-bunching in the 100~nm nanopillar is much smaller and reaches the Poissonian level in less 500~ns, indicating a high quantum efficiency. Due to the data noise in the log plot, the short time scale blinking in the 100~nm nanopillar is more clearly seen in the linear plot in Fig.~3(d) of the main text, and it is very well fitted with only one dark state.
\begin{figure}[htbp]
\includegraphics[width=0.8400\textwidth,angle=0]{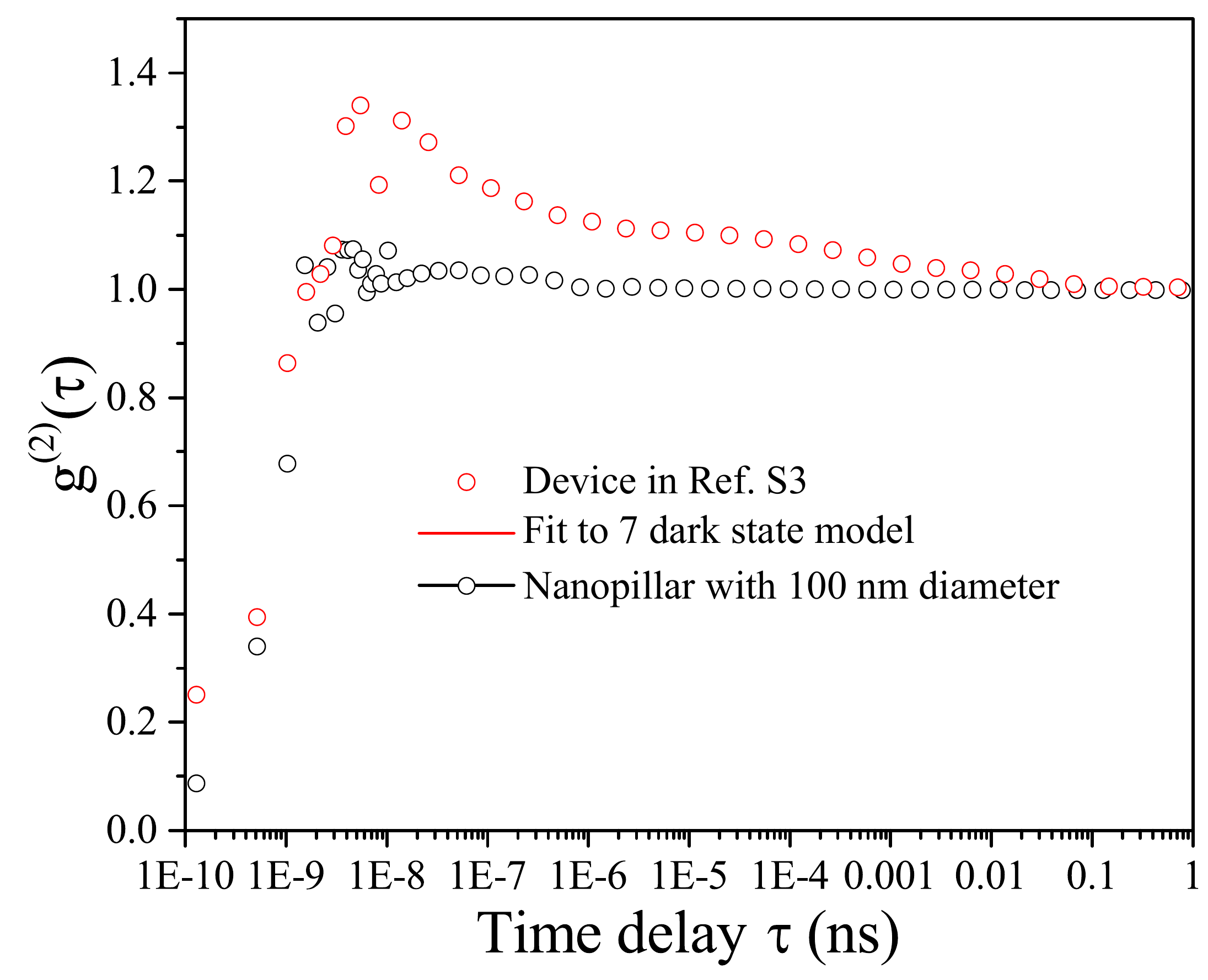}
\caption{\label{Fig2_s3.} \textbf{Quantum efficiency comparison.} Long time scale $g^{(2)}(\tau)$ data of nanopillar with 100~nm (black dots) and device 1 in  Ref.~\onlinecite{Davanco2014} (red dots), the red line is the best fit with a 7 dark states model.}
\end{figure}


\begin{thebibliography}{10}
\bibitem{NIST}
The identification of any commercial product or trade name is used to foster understanding. Such identification does not imply recommendation or endorsement or by the National Institute of Standards and Technology, nor does it imply that the materials or equipment identified are necessarily the best available for the purpose.
\end{thebibliography}
\end{document}